\documentclass[a4paper,11pt]{article}
\usepackage{graphicx}
\usepackage{jheppub}
\usepackage[T1]{fontenc} 
\usepackage{graphicx}
\usepackage{subfigure}
\usepackage[above,below]{placeins}
\usepackage{multirow}
\usepackage{longtable}
\usepackage{array}
\allowdisplaybreaks[2]
\title{\boldmath
Schwinger effect of a relativistic boson entangled with a qubit} \author[a]{  Yujie Li  }
\affiliation{Department of Physics  \&  State Key Laboratory of Surface Physics,   Fudan University,\\ Shanghai 200433, China}
\author[a]{  Qingqing Mao  }
\author[a,b,1]{Yu Shi,\note{Corresponding author.}}
\affiliation[a]{Department of Physics  \&  State Key Laboratory of Surface Physics, Fudan University,\\ Shanghai 200433, China}
\affiliation[b]{Collaborative Innovation Center of Advanced Microstructures, Fudan University, \\Shanghai 200433, China}
\emailAdd{13110190065@fudan.edu.cn}
\emailAdd{13110190010@fudan.edu.cn}
\emailAdd{yushi@fudan.edu.cn}

\abstract{
We use the concept of quantum entanglement to analyze the Schwinger effect on an entangled state of a qubit and a bosonic mode coupled with the electric field. As a consequence of the Schwinger production of particle-antiparticle pairs, the electric field decreases both the correlation and the entanglement between the qubit and the particle mode. This work exposes a profound difference between bosons and fermions. In the bosonic case,   entanglement between the qubit and the antiparticle mode cannot be caused by the  Schwinger effect on the preexisting entanglement between the qubit and the particle mode, but  correlation can.}

\begin{document}
\maketitle
\flushbottom



\section{Introduction}

In  relativistic quantum field theory, a particle is not eternal. Known as  Schwinger effect,   in a strong electromagnetic field, the vacuum decays into particle-antiparticle  pairs~\cite{Schwinger},  and  likewise, a particle becomes a superposition state involving both particles and antiparticles.  Many experimental efforts have been made to observe Schwinger effect~\cite{rmp}, though have not yet been successful, because  the rate is very low.

In this paper, the question we address is how the correlation and  entanglement between a qubit and a bosonic particle is inherited by that between the qubit on one hand,  and the particles and   antiparticles generated by the Schwinger effect on the other. Here the qubit is a simple representation of another particle uncoupled with the electric field.

Recent years witnessed the application of the concepts and measures of quantum entanglement to various areas of quantum sciences. The measures of quantum entanglement can well characterize quantum correlations in quantum states and are independent of any observable. In field theory, we partition the system  in terms of the modes~\cite{shi}.  Quantum entanglement in the Schwinger effect of Dirac or Klein-Gordon field, between a subsystem and the rest of the system, as measured by the von Neumann entropy of the reduced density matrix, was calculated~\cite{Ebadi,gavrilov}. Pairwise correlation and entanglement were also studied for Dirac field~\cite{lidaishi},  by using mutual information  and logarithmic negativity as the measures.

Pairwise correlation and entanglement are between two parts A and B, the combination of which  is described in terms of  the density matrix $\rho_{AB}$, obtained by tracing out other parts sharing a pure state. $\rho_{AB}$  usually represents a mixed state, with pure state  a special case. The reduced density matrix of A is  $\rho_A \equiv Tr_B(\rho_{AB})$, similarly,  $\rho_B \equiv Tr_A(\rho_{AB})$. The mutual information  in   $\rho_{AB}$ is then~\cite{nielsen}
\begin{equation}
I(\rho_{AB}) = S(\rho_A) +S(\rho_B) - S(\rho _{AB}), \end{equation}
where $S(\rho)\equiv -Tr(\rho\log_2\rho)$ is the von Neumann entropy of  $\rho$.   $I(\rho _{AB})=0$ if $\rho _{AB}$ is a product pure state. Hence  $I(\rho_{AB})$ measures  a kind of   distance  from  a product pure state, containing  both quantum entanglement and classical correlation. The logarithmic negativity  $N(\rho_{AB} )$, which is a measure of the quantum entanglement between A and B in $\rho_{AB} $, is defined as~\cite{vidal}
\begin{equation}
N(\rho_{AB} ) \equiv \log _2 \| \rho_{AB}^{T_A} \| , \end{equation}
where $ \| \rho_{AB} ^{T_A} \|$ is the sum of the absolute values of the eigenvalues of  the partial transpose $\rho ^{T_A}$  of the original density matrix $\rho_{AB}$  with respect to  subsystem   A. The partial transpose can also be made with respect to B, without changing the result of $N(\rho_{AB} )$.

In this paper, we consider the Schwinger effect of a one-boson state, which transforms the one-boson state to a superposition of different number states of the particle and antiparticle modes. We study  pairwise mutual information and  quantum entanglement between a qubit and the particle mode,  and that between the qubit and the  antiparticle mode.  An introduction to Schwinger effect is made in Sec.~\ref{se}. The quantum state is described in Sec.~\ref{state}. The correlation between the qubit and the particle mode $\mathbf{q}$ is  calculated in Sec.~\ref{p,q}. The correlation between the qubit and the antiparticle mode $\mathbf{-q}$ is  calculated in Sec.~\ref{p,-q}. Then the effect of a pulsed electric field is discussed in  Sec.~\ref{pulse}. A summary is made in Sec.~\ref{summary}.

\section{Schwinger effect in a constant electric field \label{se}  }

Consider a scalar field $\phi(t,x)$ describing the bosons of mass $m$ and charge $q$, coupled with an electric field $E_0$ along $z$ direction, satisfying the Klein-Gordon equation in the four-dimensional Minkowski space
\begin{equation} [( \partial _\mu  - i e A_\mu )( \partial^\mu  - i e A^\mu) + m^2 ]\phi (t,x) = 0 , \end{equation}
where $A_\mu  = (0,0,0, - E_0 t)$, $\phi (t,x)$ is the scalar field, which can be expanded in terms of the mode functions as  $\phi (t,x) = \sum_{\mathbf{k}} (  a_{\mathbf{k}} \phi_{\mathbf{k}}(t,x)+ b_{\mathbf{k}}^\dagger  \phi^{*}_{\mathbf{k}}(t,x)),$
where $\mathbf{k}$ denotes the momentum,  $a_{\mathbf{k}}$ is the annihilation operator of the particle, while $b_{\mathbf{k}}^\dagger $ is the creation operator of the antiparticle.

The Bogoliubov transformation between the in and the out modes, for $t_{in}=-\infty$ and $t_{out}=+\infty$ respectively,  is
\begin{equation}
\phi^\mathrm{in}_{\mathbf{k}} = \alpha_{ \mathbf{k} } \phi^\mathrm{out}_{\mathbf{k}} + \beta_{ \mathbf{k} } \phi^\mathrm{out *}_{\mathbf{-k}},
\end{equation}
where $\alpha_{\mathbf{k}}$ and $\beta_{\mathbf{k}}$ are Bogoliubov coefficients~\cite{brout,kim,Ebadi}
\begin{equation} \alpha_{\mathbf{k}} = \frac{\sqrt{2\pi}}{\Gamma (-\nu)} e^{\frac{-i \pi (\nu +1 )}{2}} ,\ \ \beta_{\mathbf{k}}= e^{-i\pi \nu} ,   \end{equation}
with $\nu = -\frac{1}{2}-i\frac{\mu}{2}$, $\mu = \frac{k^2_\bot + m^2}{eE_0}$,
satisfying $|\alpha_{\mathbf{k}} |^2 -  | \beta_{\mathbf{k}} |^2  = 1$.   The corresponding annihilation and creation operators of the in and the out modes are related as
\begin{equation} a^{\mathrm{in}}_{\mathbf{k}} = \alpha^*_{\mathbf{k}} a^{\mathrm{out}}_{\mathbf{k}} - \beta^*_{\mathbf{k}} b^{\mathrm{out}\dag}_{\mathbf{-k}} , \end{equation}
\begin{equation} b^{\mathrm{in}}_{\mathbf{k}} = \alpha^*_{\mathbf{k}} b^{\mathrm{out}}_{\mathbf{k}} - \beta^*_{\mathbf{k}} a^{\mathrm{out}\dag}_{\mathbf{-k}}. \end{equation}

Consequently the in-vacuum state for each mode becomes a superposition state of the out modes~\cite{brout,Ebadi},
\begin{equation} |0_{\mathbf{k}} ,0_{\mathbf{-k}} \rangle^{\mathrm{in}} = \frac{1}{\alpha_{\mathbf{k}}} \sum_{n=0}^{\infty} \left( \frac{\beta^*_{\mathbf{k}}}{\alpha^*_{\mathbf{k}}}\right)^n |n_{\mathbf{k}} ,n_{\mathbf{-k}} \rangle^{\mathrm{out}} ,    \label{cebvacuum}  \end{equation}
where $n_{\mathbf{k}}$ is the number of particles, $n_{\mathbf{-k}}$ is the number of  antiparticles. It indicates the distribution of the created particles and antiparticles due to the Schwinger effect when an electric field is applied.

Similarly, from $|1_{\mathbf{k}} ,0_{\mathbf{-k}} \rangle^{\mathrm{in}} =  a^{\mathrm{in}\dag}_{\mathbf{k}} |0_{\mathbf{k}} ,0_{\mathbf{-k}} \rangle^{\mathrm{in}}$, one obtains
\begin{equation} |1_{\mathbf{k}} ,0_{\mathbf{-k}} \rangle^{\mathrm{in}} = \frac{1}{|\alpha_{\mathbf{k}} |^2 } \sum_{n=0}^{\infty} \left( \frac{\beta^*_{\mathbf{k}}}{\alpha^*_{\mathbf{k}}}\right)^n \sqrt{n + 1} | (n + 1 )_{\mathbf{k}} ,n_{\mathbf{-k}} \rangle^{\mathrm{out}} ,\label{cebone} \end{equation}
which indicates the distribution of the created particles and antiparticles resulting from the effect of the electric field on one-particle state. We refer to this also as the Schwinger effect.

\section{The initial entangled state      \label{state} }

Now we investigate the influence of the electric field on the state of a qubit $\sigma$  entangled with a bosonic  particle  of momentum  $\mathbf{q}$, which is an excitation of the scalar field discussed above,
\begin{equation}
|\Phi_{\sigma,\mathbf{q}}\rangle =   \varepsilon | \uparrow\rangle |0_\mathbf{q} \rangle^{\mathrm {in}}   + \sqrt{1 - \varepsilon^2 } | \downarrow\rangle |1_\mathbf{q} \rangle^{\mathrm {in}} , \label{originalstate}
\end{equation}
where $\varepsilon$ is a coefficient, the basis states of the qubit are denoted as $|\uparrow\rangle$ and $|\downarrow\rangle$. Obviously, the von Neumann entropy of the  reduced matrices $\rho_\sigma$ and $\rho_{\mathbf{q}}$ are both equal to
\begin{equation}
 S(\varepsilon)= - \varepsilon^2 \log_2 \varepsilon^2 - (1 - \varepsilon^2 ) \log_2 (1 - \varepsilon^2 ).
\end{equation}
Being a pure state, the  von Neumann entropy of $|\Phi_{\sigma,\mathbf{q}}\rangle$ is $0$, therefore the mutual information
\begin{equation} I(\Phi_{\sigma,\mathbf{q}})=2S(\varepsilon).
\end{equation}
The  entanglement entropy, characterizing the entanglement between the qubit $\sigma$ and the in mode $\mathbf{q}$ is  just $S(\varepsilon)$.

With the mode $\mathbf{-q}$ also considered, $|\Phi_{\sigma,\mathbf{q}}\rangle$   can be rewritten as
\begin{equation} |\Phi_{\sigma,\mathbf{q},\mathbf{-q} } \rangle^{\mathrm {in}} = (\varepsilon | \uparrow\rangle^{\mathrm {in}} |0_\mathbf{q} \rangle^{\mathrm {in}} + \sqrt{1 - \varepsilon^2 } | \downarrow\rangle^{\mathrm {in}} |1_\mathbf{q} \rangle^{\mathrm {in}}) |0_\mathbf{-q} \rangle^{\mathrm {in}} .  \label{initial}
\end{equation}
Because of Bogoliubov transformation given in   Eq.(\ref{cebvacuum}),  one  obtains
\begin{align}
&| \Phi_{\sigma,\mathbf{q},\mathbf{-q}} \rangle^{\mathrm {in}}  \nonumber  \\
&=   \frac{ \varepsilon  }{\alpha_\mathbf{q} } \sum _{ n=0 }^\infty   \frac{\beta^{*n}_\mathbf{q} }{\alpha^{*n}_\mathbf{q} }  | \uparrow, n_\mathbf{q},n_\mathbf{-q} \rangle^{\mathrm {out}}
 + \frac{ \sqrt{1 - \varepsilon^2 } }{ | \alpha_\mathbf{q}|^2 } \sum _{ n=0 }^\infty    \frac{\beta^{*n}_\mathbf{q} }{\alpha^{*n}_\mathbf{q} }   \sqrt{n+ 1}|\downarrow, (n +1)_\mathbf{q},n_\mathbf{-q} \rangle^{\mathrm {out}}.
\end{align}
The density matrix $\rho_{\sigma,\mathbf{q},\mathbf{-q} } =  | \Phi_{\sigma,\mathbf{q},\mathbf{-q}} \rangle^{\mathrm {in}}  {^{\mathrm {in}}}{\langle \Phi_{\sigma,\mathbf{q},\mathbf{-q}} |} $  is thus
\begin{align}
&\rho_{\sigma,\mathbf{q},\mathbf{-q} }   \nonumber  \\
& = \frac{ \varepsilon^2 }{ |\alpha_\mathbf{q}|^2 } \sum _{ n,m=0}^\infty   \frac{\beta^{*n}_\mathbf{q} \beta^m_\mathbf{q} }{\alpha^{*n}_\mathbf{q}\alpha^m_\mathbf{q} }   |\uparrow, n_\mathbf{q} , n_\mathbf{-q} \rangle  \langle \uparrow, m_\mathbf{q} , m_\mathbf{-q} |   \nonumber  \\
& +  \frac{ \varepsilon \sqrt{ 1 - \varepsilon^2 }  }{ |\alpha_\mathbf{q}|^2 \alpha_\mathbf{q} } \sum _{  n,m=0}^\infty   \frac{\beta^{*n}_\mathbf{q} \beta^m_\mathbf{q} }{\alpha^{*n}_\mathbf{q}\alpha^m_\mathbf{q} }  \sqrt{m + 1}| \uparrow, n_\mathbf{q} , n_\mathbf{-q} \rangle \langle \downarrow, (m+1)_\mathbf{q} , m_\mathbf{-q} |   \nonumber  \\
& +  \frac{ \varepsilon \sqrt{ 1 - \varepsilon^2 }  }{ |\alpha_\mathbf{q}|^2 \alpha^*_\mathbf{q} } \sum _{n,m=0}^\infty   \frac{\beta^{*n}_\mathbf{q} \beta^m_\mathbf{q} }{\alpha^{*n}_\mathbf{q}\alpha^m_\mathbf{q} }  \sqrt{n + 1}| \downarrow, (n + 1)_\mathbf{q} , n_\mathbf{-q} \rangle  \langle \uparrow, m_\mathbf{q} , m_\mathbf{-q} |   \nonumber  \\
& + \frac{ 1 - \varepsilon^2 }{ |\alpha_\mathbf{q}|^4 } \sum _{ n,m=0  }^\infty   \frac{\beta^{*n}_\mathbf{q} \beta^m_\mathbf{q} }{\alpha^{*n}_\mathbf{q}\alpha^m_\mathbf{q} }   \sqrt{(n + 1) (m + 1)}| \downarrow, (n + 1)_\mathbf{q} , n_\mathbf{-q} \rangle  \langle \downarrow, (m + 1)_\mathbf{q} , m_\mathbf{-q} |,
\end{align}
which indicates that the Bogoliubov transformation causes in  mode $\mathbf{q}$ to be replaced by the  out modes $\mathbf{q}$  and $\mathbf{-q}$. How the original correlation and entanglement are inherited between the qubit and these out modes will be investigated below.    For brevity, we have  omitted the superscript ``out''.

\section{Correlation and entanglement between the qubit   $   \sigma $ and the mode  $\mathbf{q}$  \label{p,q}    }

We first study the correlation and entanglement between the qubit $\sigma$ and the out mode $\mathbf{q} $. Tracing out the mode $  -\mathbf{q} $, we obtain the reduced density matrix of  the  qubit $\sigma$ and $\mathbf{q} $, $\rho _{\sigma,\mathbf{q}}  =  {\mathrm{Tr}}_{ -\mathbf{q}}( \rho_{\sigma,\mathbf{q},\mathbf{-q} } )$ as
\begin{align}
\rho_{\sigma,\mathbf{q}  } = & \frac{\varepsilon^2 }{ |\alpha_\mathbf{q}|^2 } \sum _{  n=0  }^\infty   \left| \frac{\beta_\mathbf{q} }{\alpha_\mathbf{q} } \right|^{2n} | \uparrow, n_\mathbf{q}   \rangle  \langle \uparrow, n_\mathbf{q}   |   \nonumber  \\
& +  \frac{ \varepsilon \sqrt{ 1 - \varepsilon^2 }  }{ |\alpha_\mathbf{q}|^2 \alpha_\mathbf{q} } \sum _{  n=0  }^\infty   \left| \frac{\beta_\mathbf{q} }{\alpha_\mathbf{q} } \right|^{2n} \sqrt{n + 1}|  \uparrow, n_\mathbf{q}   \rangle  \langle \downarrow, (n +1)_\mathbf{q}   |   \nonumber  \\
& +  \frac{ \varepsilon \sqrt{ 1 - \varepsilon^2 }  }{ |\alpha_\mathbf{q}|^2 \alpha^*_\mathbf{q} } \sum _{  n=0  }^\infty   \left| \frac{\beta_\mathbf{q} }{\alpha_\mathbf{q} } \right|^{2n} \sqrt{n + 1}|  \downarrow, (n + 1)_\mathbf{q}   \rangle  \langle \uparrow, n_\mathbf{q}   |   \nonumber  \\
& + \frac{ 1 - \varepsilon^2 }{ |\alpha_\mathbf{q}|^4 } \sum _{  n=0  }^\infty   \left| \frac{\beta_\mathbf{q} }{\alpha_\mathbf{q} } \right|^{2n}  (n + 1)  |  \downarrow, (n + 1)_\mathbf{q}  \rangle  \langle  \downarrow, (n + 1)_\mathbf{q}   |.
\end{align}
With the above summation expression,    in the subspace of  $ \{ | \uparrow, n_\mathbf{q} \rangle  , |  \downarrow, (n + 1)_\mathbf{q} \rangle \} $, $(n =0, 1, 2, \cdots)$, $\rho_{\sigma,\mathbf{q}  } $  is a block matrix, with non-zero eigenvalues
\begin{equation}  \frac{1}{| \alpha_\mathbf{q}|^2 } \left| \frac{\beta_\mathbf{q} }{\alpha_\mathbf{q} } \right|^{2n} \left[ \varepsilon^2 +  \frac{ ( n + 1) (1 -  \varepsilon^2 ) }{| \alpha_\mathbf{q}|^2 } \right].   \end{equation}

Tracing out the field mode $\mathbf{q}$ in $\rho_{\sigma,\mathbf{q}}$ yields $\rho_\sigma  = {\mathrm{Tr}}_\mathbf{q}(\rho_{\sigma,\mathbf{q}})$, which is
\begin{equation}
\rho_\sigma  =  \varepsilon^2 | \uparrow  \rangle \langle \uparrow  |  + ( 1 - \varepsilon^2 ) | \downarrow  \rangle \langle \downarrow  | ,  \end{equation}
with eigenvalues $\varepsilon^2$, $ 1 - \varepsilon^2 $. This remains unchanged from the reduced density matrix of qubit $\sigma$ obtained from $
|\Phi_{\sigma,\mathbf{q}}\rangle $ in (\ref{originalstate}), as nothing is done on the  qubit $\sigma$.

Tracing out the qubit  $\sigma$ in $\rho_{\sigma,\mathbf{q}}$ yields  $\rho_\mathbf{q} =  \mathrm{Tr}_\sigma \rho(\sigma,\mathbf{q}) $,  which is
\begin{equation}
\rho_\mathbf{q} = \frac{1}{ |\alpha_\mathbf{q}|^2 } \sum _{  n=0  }^\infty   \left| \frac{\beta_\mathbf{q} }{\alpha_\mathbf{q} } \right|^{2n} \left[ \varepsilon^2 +  \frac{ n (1 - \varepsilon^2 ) }{ |\beta_\mathbf{q}|^2 } \right]  | n_ \mathbf{q} \rangle \langle n_ \mathbf{q} | ,  \end{equation}
with eigenvalues
\begin{equation}  \frac{1}{| \alpha_\mathbf{q}|^2 } \left| \frac{\beta_\mathbf{q} }{\alpha_\mathbf{q} } \right|^{2n} \left[ \varepsilon^2 +  \frac{   n  (1 -  \varepsilon^2 ) }{| \beta_\mathbf{q}|^2 } \right], \ \ n = 0,1,2,\cdots\cdots   \end{equation}

Then we obtain the mutual information $ I( \rho_{\sigma ,\mathbf{q}} )= S(\rho_{\sigma  }) + S(\rho_{ \mathbf{q}}) - S(\rho_{\sigma ,\mathbf{q}}) $ as
\begin{align}
& I( \rho_{\sigma ,\mathbf{q}} )   \nonumber  \\
& =  - \varepsilon^2 \log_2 \varepsilon^2 - (1 - \varepsilon^2 ) \log_2 (1 - \varepsilon^2 )  \nonumber  \\
& - \sum _{  n=0  }^\infty  \frac{| \beta_\mathbf{q} |^{2n} }{| \alpha_\mathbf{q}|^{2(n + 1 )} }   \left[ \varepsilon^2 +  \frac{   n  (1 - \varepsilon^2 ) }{| \beta_\mathbf{q}|^2 } \right] \log_2 \left[  \frac{| \beta_\mathbf{q} |^{2n} }{| \alpha_\mathbf{q}|^{2(n + 1 )} }  \left( \varepsilon^2 +  \frac{   n  (1 -  \varepsilon^2 ) }{| \beta_\mathbf{q}|^2 } \right)  \right]   \nonumber  \\
& +  \sum _{  n=0  }^\infty  \frac{| \beta_\mathbf{q} |^{2n} }{| \alpha_\mathbf{q}|^{2(n + 1 )} }  \left[ \varepsilon^2 +  \frac{ ( n +1 )  (1 - \varepsilon^2 ) }{| \alpha_\mathbf{q}|^2 } \right] \log_2 \left[  \frac{| \beta_\mathbf{q} |^{2n} }{| \alpha_\mathbf{q}|^{2(n + 1 )} }  \left( \varepsilon^2 +  \frac{ ( n +1 ) (1 -  \varepsilon^2 ) }{| \alpha_\mathbf{q}|^2 } \right)  \right],
\end{align}
which depends on the coefficient parameter $ \varepsilon $ and  the strength of the electric field $ E_0 $. When $ E_0 = 0 $,  $ I( \rho_{\sigma ,\mathbf{q}} )$ reduces to $S(\varepsilon)$.

\begin{figure}
\centering
\scalebox{0.7}{\includegraphics{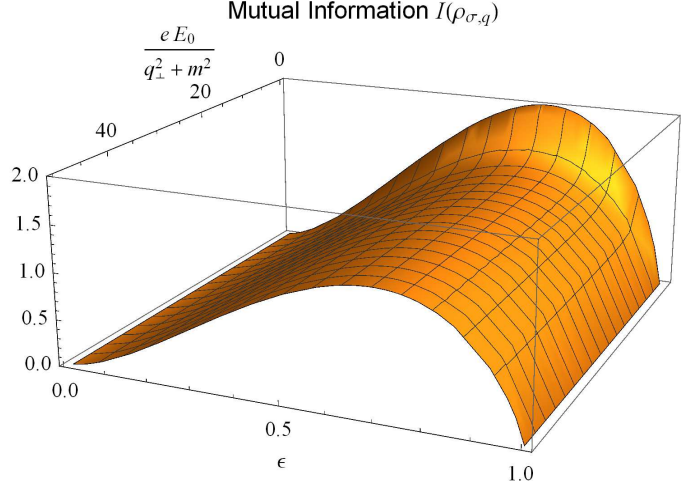}}
\caption{  The mutual information $I(\rho_{\sigma ,\mathbf{q}})$  as a function of the dimensionless parameters $\frac{eE_0}{q_\perp^2+m^2}$ and $\varepsilon $, where $E_0$ is the strength of the constant electric field, and $\varepsilon $ is the coefficient parameter of the initial entangled state.    \label{Fig1} }
\end{figure}

The dependence of the mutual information $I(\rho_{\sigma ,\mathbf{q}})$  on the electric field $E_0$ and the parameter $\varepsilon $ is shown in Fig.~\ref{Fig1}. For a fixed value of $\varepsilon $, $I(\rho_{\sigma ,\mathbf{q}})$ monotonically decreases with the increase of the electric field $E_0$ and asymptotically approaches a certain nonvanishing value independent of $E_0$. For $\varepsilon $ less or larger than $\frac{1}{\sqrt{2}} $, the closer to $\frac{1}{\sqrt{2}} $ the parameter $\varepsilon $ is, the quicker $I(\rho_{\sigma ,\mathbf{q}})$ decreases with the increase of $E_0$ when $E_0$ is small. For any given value of $E_0$ and for $\varepsilon $ less or larger than $\frac{1}{\sqrt{2}} $, the farther to $\frac{1}{\sqrt{2}} $ the parameter $\varepsilon $ is, the smaller $I(\rho_{\sigma ,\mathbf{q}})$ is. The mutual information $I(\rho_{\sigma ,\mathbf{q}})$ becomes zero as $\varepsilon = 0 $ or $1$, in  which  case the mutual information vanishes even in the absence of the electric field.

We  use logarithmic negativity to measure the entanglement.
After making the partial transpose of the density matrix $\rho_{\sigma ,\mathbf{q}  }$ with respect to  $ \sigma  $, we obtain
\begin{align}
\rho^{\mathrm{T}_\sigma  }_{ \sigma ,\mathbf{q}  }  = & \frac{\varepsilon^2 }{ |\alpha_\mathbf{q}|^2 } \sum _{  n=0  }^\infty   \left| \frac{\beta_\mathbf{q} }{\alpha_\mathbf{q} } \right|^{2n} |  \uparrow, n_\mathbf{q}   \rangle  \langle \uparrow, n_\mathbf{q}   |   \nonumber  \\
& +  \frac{ \varepsilon \sqrt{ 1 - \varepsilon^2 }  }{ |\alpha_\mathbf{q}|^2 \alpha_\mathbf{q} } \sum _{  n=0  }^\infty   \left| \frac{\beta_\mathbf{q} }{\alpha_\mathbf{q} } \right|^{2n} \sqrt{n + 1}|  \downarrow, n_\mathbf{q}   \rangle  \langle \uparrow, (n +1)_\mathbf{q}   |   \nonumber  \\
& +  \frac{ \varepsilon \sqrt{ 1 - \varepsilon^2 }  }{ |\alpha_\mathbf{q}|^2 \alpha^*_\mathbf{q} } \sum _{  n=0  }^\infty   \left| \frac{\beta_\mathbf{q} }{\alpha_\mathbf{q} } \right|^{2n} \sqrt{n + 1}|  \uparrow, (n + 1)_\mathbf{q}   \rangle  \langle \downarrow, n_\mathbf{q}   |   \nonumber  \\
& + \frac{ 1 - \varepsilon^2 }{ |\alpha_\mathbf{q}|^4 } \sum _{  n=0  }^\infty   \left| \frac{\beta_\mathbf{q} }{\alpha_\mathbf{q} } \right|^{2n}  (n + 1)  |  \downarrow, (n + 1)_\mathbf{q}  \rangle  \langle  \downarrow, (n + 1)_\mathbf{q}   | ,
\end{align}
which is a block matrix in   the subspace of  $\{  |  \uparrow, (n + 1)_\mathbf{q} \rangle  , |  \downarrow, n_\mathbf{q}  \rangle \} $, $(n = 0,1,2,\cdots) $. Therefore the eigenvalues of $ \rho^{\mathrm{T}_\sigma  }_{\sigma ,\mathbf{q}  } $ are
\begin{equation*}   \frac{ \varepsilon^2 }{ |\alpha_\mathbf{q}|^2 } ,  \end{equation*}
\begin{align}  & \ \ \ \ \ \ \ \frac{ 1 }{ 2 |\alpha_\mathbf{q}|^2  }  \left| \frac{\beta_\mathbf{q} }{\alpha_\mathbf{q} } \right|^{2n}  \left[ \   \left| \frac{\beta_\mathbf{q} }{\alpha_\mathbf{q} } \right|^2 \varepsilon^2 + \frac{ n ( 1 - \varepsilon^2 )} { |\beta_\mathbf{q}|^2 } \right. \pm  \nonumber \\
& \left. \sqrt{ \left(  \left| \frac{\beta_\mathbf{q} }{\alpha_\mathbf{q} } \right|^2 \varepsilon^2 + \frac{ n ( 1 - \varepsilon^2 )} { |\beta_\mathbf{q}|^2 } \right )^2 +  \frac{ 4 \varepsilon^2 ( 1 - \varepsilon^2 )} { |\alpha_\mathbf{q}|^2 } }  \ \right], \ \ n = 0,1,2,\cdots\cdots
\end{align}

Thus the logarithmic negativity is
\begin{align}
&N( \rho _{\sigma ,\mathbf{q} } )    \nonumber \\
&= \log_2 \left[  \frac{ \varepsilon^2 }{ |\alpha_\mathbf{q}|^2 } + \sum_{n=0}^\infty \frac{| \beta_\mathbf{q} |^{2n} }{| \alpha_\mathbf{q}|^{2(n + 1 )} }  \sqrt{ \left(  \left| \frac{\beta_\mathbf{q} }{\alpha_\mathbf{q} } \right|^2 \varepsilon^2 + \frac{ n ( 1 - \varepsilon^2 )} { |\beta_\mathbf{q}|^2 } \right )^2 +  \frac{ 4 \varepsilon^2 ( 1 - \varepsilon^2 )} { |\alpha_\mathbf{q}|^2 } } \ \right] ,  \end{align}
which describes the quantum entanglement between  $  \sigma  $ and $ \mathbf{q} $.  $ N( \rho _{\sigma ,\mathbf{q} } ) =\log_2 [ 1 + 2 \varepsilon \sqrt{1 - \varepsilon^2 } ] $  when  $ E_0 = 0 $.

\begin{figure}
\centering
\scalebox{0.7}{\includegraphics{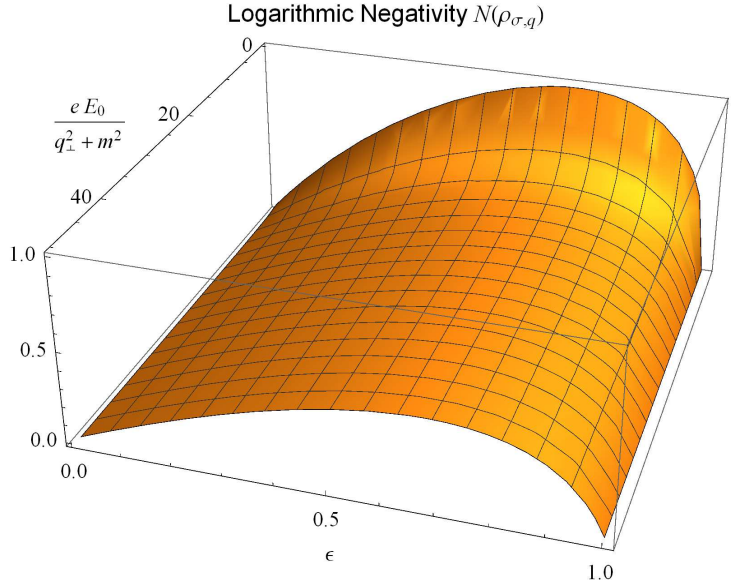}}
\caption{  The logarithmic negativity $N(\rho_{\sigma,\mathbf{q}})$ as a function of the dimensionless parameters $\frac{eE_0}{q_\perp^2+m^2}$ and $\varepsilon $, where $E_0$ is the strength of the constant electric field, and $\varepsilon $ is the coefficient parameter of the initial entangled state. \label{Fig2} }
\end{figure}

Fig.~\ref{Fig2} shows how the logarithmic negativity $N(\rho_{\sigma ,\mathbf{q}})$   depends on the strength of the electric field $ E_0$ and the parameter $\varepsilon $. The variation trend of $N(\rho_{\sigma ,\mathbf{q}})$ with respect to $ E_0$ and $\varepsilon $ is similar to that of $I(\rho_{\sigma ,\mathbf{q}})$. But                           when $E_0$ is small, $N(\rho_{\sigma ,\mathbf{q}})$ decreases more rapidly with the increase of $E_0$  than   $I(\rho_{\sigma ,\mathbf{q}})$ does, indicating that the entanglement  is more sensitive to the coupling with the electric field. Like $I(\rho_{\sigma ,\mathbf{q}})$,  $N(\rho_{\sigma ,\mathbf{q}})$ monotonically decreases with the increase of the electric field $E_0$ and approaches a certain nonzero asymptotic value as $ E_0 \rightarrow   \infty $.

The Schwinger effect of two entangled fermions of momenta  $\mathbf{p}$ and  $\mathbf{q}$~\cite{lidaishi}  can reduce effectively to a fermion counterpart of our present  bosonic  problem, under the constraint that the electric field does not couple the mode $\mathbf{p}$, thus fixing the Bogoliubov coefficients of the $\mathbf{p}$ mode to be $\alpha_{\mathbf{p}}=0$, $\beta_{\mathbf{p}}=1$, thereby  reducing  mode $\mathbf{p}$  to our  qubit $\sigma$.  With the increase of $E_0$, the monotonic  decrease towards $0$ of the mutual information and entanglement between fermionic mode $\mathbf{p}$ and mode $\mathbf{q}$ reduces to  those between the qubit $\mathbf{\sigma}$ and fermionic mode $\mathbf{q}$ here. The boson-fermion comparison will be discussed in the summary.

It is also interesting to make comparison with the bosons in Unruh effect~\cite{fuentes1,Eduardo,Pan} and near a dilaton black hole~\cite{Jieci}, with the role of the electric field in our case   replaced as the acceleration, but  the Bogoliubov coefficient  that is the counterpart of $|\beta_\mathbf{k}|^2 $ can be  arbitrarily large,  making  the entanglement  disappear in the limiting  cases.   In contrast, in our present case, $|\beta_\mathbf{k} |^2   < 1  $, consequently the  entanglement in  $\rho_{\sigma ,\mathbf{q}}$  persists as $ E_0 \rightarrow   \infty $.

\section{Correlation and entanglement between the qubit   $   \sigma $ and the mode  $\mathbf{-q}$     \label{p,-q}     }

Now  we  study the correlation and the entanglement between   $ \sigma $ and $\mathbf{-q}$. Tracing out the  mode $ \mathbf{q} $, we obtain the reduced density matrix of  $ \sigma $ and $\mathbf{-q}$, $\rho _{\sigma ,\mathbf{-q}}  =  {\mathrm{Tr}}_{ \mathbf{q}}( \rho_{\sigma ,\mathbf{q},\mathbf{-q} } )$,  as
\begin{align}
\rho_{\sigma ,\mathbf{-q}  } = & \frac{\varepsilon^2 }{ |\alpha_\mathbf{q}|^2 } \sum _{  n=0  }^\infty   \left| \frac{\beta_\mathbf{q} }{\alpha_\mathbf{q} } \right|^{2n} | \uparrow, n_\mathbf{-q}   \rangle  \langle  \uparrow, n_\mathbf{-q}   |   \nonumber  \\
& +  \frac{ \varepsilon \sqrt{ 1 - \varepsilon^2 } \beta^*_\mathbf{q}  }{ |\alpha_\mathbf{q}|^4  } \sum _{  n=0  }^\infty   \left| \frac{\beta_\mathbf{q} }{\alpha_\mathbf{q} } \right|^{2n} \sqrt{n + 1}|  \uparrow, (n+ 1)_\mathbf{-q}   \rangle  \langle  \downarrow,  n _\mathbf{-q}   |   \nonumber  \\
& +  \frac{ \varepsilon \sqrt{ 1 - \varepsilon^2  } \beta _\mathbf{q} }{ |\alpha_\mathbf{q}|^4  } \sum _{  n=0  }^\infty   \left| \frac{\beta_\mathbf{q} }{\alpha_\mathbf{q} } \right|^{2n} \sqrt{n + 1}|  \downarrow,  n _\mathbf{-q}   \rangle  \langle \uparrow, (n+ 1)_\mathbf{-q}   |   \nonumber  \\
& + \frac{ 1 - \varepsilon^2 }{ |\alpha_\mathbf{q}|^4 } \sum _{  n=0  }^\infty   \left| \frac{\beta_\mathbf{q} }{\alpha_\mathbf{q} } \right|^{2n}  (n + 1)  |  \downarrow,  n _\mathbf{-q}  \rangle  \langle  \downarrow,  n _\mathbf{-q}   | ,
\end{align}
which is a block matrix in the subspace of  $\{ |  \uparrow,  n_\mathbf{q} \rangle  , |  \downarrow, (n - 1 )_\mathbf{q}  \rangle \} $,  $(n = 1,2,3,\cdots)$, thus the non-zero eigenvalues of $ \rho_{\sigma ,\mathbf{-q}  } $ are
\begin{equation}  \frac{1}{| \alpha_\mathbf{q}|^2 } \left| \frac{\beta_\mathbf{q} }{\alpha_\mathbf{q} } \right|^{2n} \left[ \varepsilon^2 +  \frac{  n (1 -  \varepsilon^2 ) }{| \beta_\mathbf{q}|^2 } \right], \ \ n = 0,1,2,\cdots. \end{equation}

Tracing out the field mode $\mathbf{-q}$ in $\rho_{\sigma ,\mathbf{-q}}$ yields $\rho_\sigma  = {\mathrm{Tr}}_\mathbf{-q}(\rho_{\sigma ,\mathbf{-q}})$,  which is
\begin{equation}
\rho_\sigma  =  \varepsilon^2 | 0_  p  \rangle \langle 0_\sigma  |  + ( 1 - \varepsilon^2 ) | 1_  p  \rangle \langle 1_\sigma  | ,  \end{equation}
with eigenvalues $\varepsilon^2$, $ 1 - \varepsilon^2 $.

Tracing out   $ \sigma $ in $\rho_{\sigma ,\mathbf{-q}}$ yields $\rho_\mathbf{-q} = {\mathrm{Tr}}_\sigma (\rho_{\sigma ,\mathbf{-q}})$,  which is
\begin{equation} \rho_\mathbf{-q} = \frac{1}{ |\alpha_\mathbf{q}|^2 } \sum _{  n=0  }^\infty   \left| \frac{\beta_\mathbf{q} }{\alpha_\mathbf{q} } \right|^{2n} \left[ \varepsilon^2 +  \frac{ (n + 1) (1 - \varepsilon^2 ) }{ |\alpha_\mathbf{q}|^2 } \right]  | n_ \mathbf{-q} \rangle \langle n_ \mathbf{-q} | ,  \end{equation}
with eigenvalues
\begin{equation}  \frac{1}{| \alpha_\mathbf{q}|^2 } \left| \frac{\beta_\mathbf{q} }{\alpha_\mathbf{q} } \right|^{2n} \left[ \varepsilon^2 +  \frac{ (n + 1) (1 - \varepsilon^2 ) }{ |\alpha_\mathbf{q}|^2 } \right],  \ \ n = 0,1,2,\cdots. \end{equation}

According to the definition of the mutual information between modes $ \sigma $ and $\mathbf{-q}$, we have
\begin{align}
& I(\rho_{\sigma ,\mathbf{-q}} )    \nonumber  \\
& =  - \varepsilon^2 \log_2 \varepsilon^2 - (1 - \varepsilon^2 ) \log_2 (1 - \varepsilon^2 )  \nonumber  \\
& +  \sum _{  n=0  }^\infty  \frac{| \beta_\mathbf{q} |^{2n} }{| \alpha_\mathbf{q}|^{2(n + 1 )} }  \left[ \varepsilon^2 +  \frac{   n (1 - \varepsilon^2 ) }{| \beta_\mathbf{q}|^2 } \right] \log_2 \left[  \frac{| \beta_\mathbf{q} |^{2n} }{| \alpha_\mathbf{q}|^{2(n + 1 )} }  \left( \varepsilon^2 + \frac{   n (1 - \varepsilon^2 ) }{| \beta_\mathbf{q}|^2 }  \right)  \right]      \nonumber  \\
& - \sum _{  n=0  }^\infty  \frac{| \beta_\mathbf{q} |^{2n} }{| \alpha_\mathbf{q}|^{2(n + 1 )} }   \left[ \varepsilon^2 +  \frac{ (n + 1)  (1 - \varepsilon^2 ) }{| \alpha_\mathbf{q}|^2 } \right] \log_2 \left[  \frac{| \beta_\mathbf{q} |^{2n} }{| \alpha_\mathbf{q}|^{2(n + 1 )} }  \left( \varepsilon^2 +  \frac{ (n + 1)  (1 - \varepsilon^2 ) }{| \alpha_\mathbf{q}|^2 }  \right)  \right] .
\end{align}

\begin{figure}
\centering
\scalebox{0.7}{\includegraphics{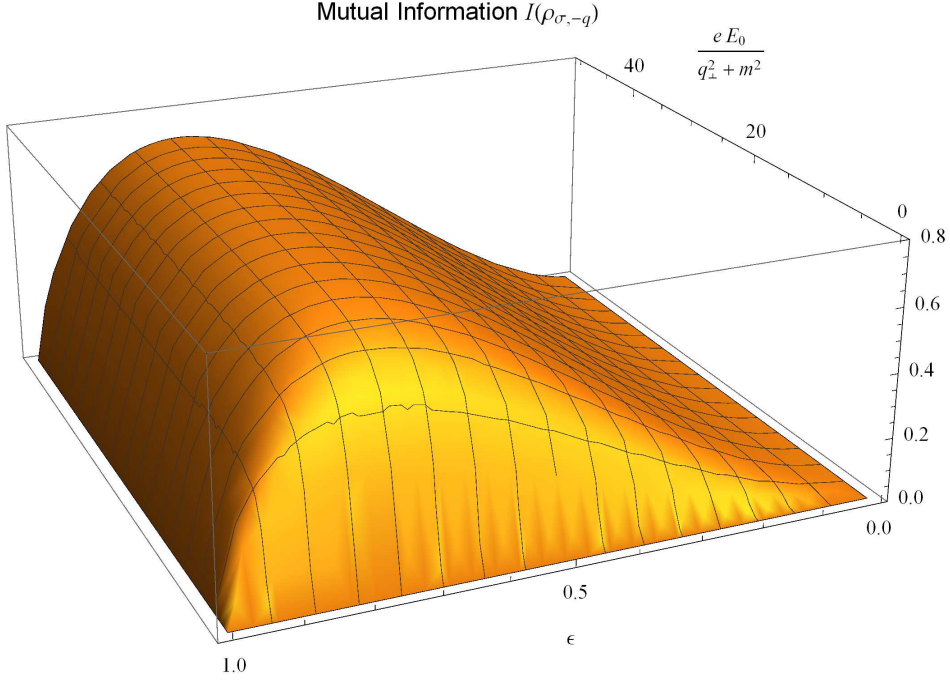}}
\caption{  The mutual information $I(\rho_{\sigma ,\mathbf{-q}})$ as a function of the dimensionless parameters $\frac{eE_0}{q_\perp^2+m^2}$ and $\varepsilon $, where $E_0$ is the strength of the constant electric field, and $\varepsilon $ is the coefficient parameter of the initial entangled state.   \label{Fig3} }
\end{figure}

The dependence of the mutual information $I(\rho_{\sigma ,\mathbf{-q}})$   on the electric field $E_0$ and the parameter $\varepsilon $ is shown in Fig.~\ref{Fig3}. $I(\rho_{\sigma ,\mathbf{-q}})$ monotonically increases with the increase of the electric field $E_0$, and asymptotically approaches a certain value independent of $E_0$. For $\varepsilon $ larger or smaller than $\frac{1}{\sqrt{2} }$, the closer to $\frac{1}{\sqrt{2} }$ the parameter $\varepsilon $ is, the larger the asymptotic value is. Moreover, when $E_0$ is small, for $\varepsilon $ larger or smaller than $\frac{1}{\sqrt{2} }$, the closer to $\frac{1}{\sqrt{2} }$ the parameter $\varepsilon $ is, the quicker $I(\rho_{\sigma ,\mathbf{-q}})$ increases with the increase of $E_0$, and the larger the value of $I(\rho_{\sigma ,\mathbf{-q}})$ is.

Now we  calculate the logarithmic negativity of $\rho_{\sigma ,\mathbf{-q}  }$. After making the partial transpose of the density matrix $\rho_{\sigma ,\mathbf{-q}  }$ with respect to the mode $  p  $, one obtains
\begin{align}
\rho^{\mathrm{T}_\sigma  }_{\sigma ,\mathbf{-q}  }  = & \frac{\varepsilon^2 }{ |\alpha_\mathbf{q}|^2 } \sum _{  n=0  }^\infty   \left| \frac{\beta_\mathbf{q} }{\alpha_\mathbf{q} } \right|^{2n} |  \uparrow, n_\mathbf{-q}   \rangle  \langle \uparrow, n_\mathbf{-q}   |   \nonumber  \\
& +  \frac{ \varepsilon \sqrt{ 1 - \varepsilon^2 } \beta^*_\mathbf{q}  }{ |\alpha_\mathbf{q}|^4  } \sum _{  n=0  }^\infty   \left| \frac{\beta_\mathbf{q} }{\alpha_\mathbf{q} } \right|^{2n} \sqrt{n + 1}|  \downarrow, (n+ 1)_\mathbf{-q}   \rangle  \langle  \uparrow,  n _\mathbf{-q}   |   \nonumber  \\
& +  \frac{ \varepsilon \sqrt{ 1 - \varepsilon^2  } \beta _\mathbf{q} }{ |\alpha_\mathbf{q}|^4  } \sum _{  n=0  }^\infty   \left| \frac{\beta_\mathbf{q} }{\alpha_\mathbf{q} } \right|^{2n} \sqrt{n + 1}|  \uparrow,  n _\mathbf{-q}   \rangle  \langle \downarrow, (n+ 1)_\mathbf{-q}   |   \nonumber  \\
& + \frac{ 1 - \varepsilon^2 }{ |\alpha_\mathbf{q}|^4 } \sum _{  n=0  }^\infty   \left| \frac{\beta_\mathbf{q} }{\alpha_\mathbf{q} } \right|^{2n}  (n + 1)  |  \downarrow,  n _\mathbf{-q}  \rangle  \langle  \downarrow,  n _\mathbf{-q}   |,
\end{align}
which is a  block matrix in the subspace of   $ \{ |  \uparrow ,  n_\mathbf{q} \rangle  , |  \downarrow, (n + 1)_\mathbf{q}  \rangle \} $, with  eigenvalues
\begin{equation*}   \frac{  1- \varepsilon^2  }{ |\alpha_\mathbf{q}|^4 } ,  \end{equation*}
\begin{align}  & \ \ \ \ \ \ \ \frac{ 1 }{ 2 |\alpha_\mathbf{q}|^2  }  \left| \frac{\beta_\mathbf{q} }{\alpha_\mathbf{q} } \right|^{2n}  \left[ \  \varepsilon^2 + \frac{ (n + 2) ( 1 - \varepsilon^2 ) |\beta_\mathbf{q}|^2 } { |\alpha_\mathbf{q}|^4 } \right. \pm  \nonumber \\
& \left. \sqrt{ \left(  \varepsilon^2 + \frac{ (n + 2) ( 1 - \varepsilon^2 ) |\beta_\mathbf{q}|^2 } { |\alpha_\mathbf{q}|^4 } \right )^2 -  \frac{ 4 \varepsilon^2 ( 1 - \varepsilon^2 ) |\beta_\mathbf{q}|^2  } { |\alpha_\mathbf{q}|^4 } }  \ \right], \ \ n = 0,1,2,\cdots.
\end{align}

Thus the logarithmic negativity of $\rho _{\sigma ,\mathbf{-q}}$ is
\begin{align}
N(\rho_{\sigma ,\mathbf{-q} } ) & = \log_2 \left[  \frac{ 1 - \varepsilon^2 }{ |\alpha_\mathbf{q}|^4 } + \sum_{n=0}^\infty \frac{| \beta_\mathbf{q} |^{2n} }{| \alpha_\mathbf{q}|^{2(n + 1 )} } \left(  \varepsilon^2 + \frac{(n + 2)( 1 - \varepsilon^2 ) |\beta_\mathbf{q}|^2 } { |\alpha_\mathbf{q}|^4 } \right ) \right]     \nonumber \\ &  = \log_2 1 = 0,   \end{align}
which means   starting from the initial state (\ref{initial} ) with any value of the parameter $\varepsilon $, which is  entangled  between $\sigma$ and in mode $\mathbf{q}$, with the action of the electric field, the  entanglement between $\sigma$ and out  mode $\mathbf{-q}$ always vanishes.

From the expressions of the mutual information of $\rho_{\sigma ,\mathbf{q} } $ and $\rho_{\sigma ,\mathbf{-q} } $, we obtain
\begin{equation}  I(\rho_{\sigma ,\mathbf{ q} } ) + I( \rho_{\sigma ,\mathbf{-q} }) = -2[ \varepsilon^2 \log_2 \varepsilon^2 + (1 - \varepsilon^2 ) \log_2 (1 - \varepsilon^2 ) ], \label{compl}   \end{equation}
implying that Schwinger effect redistributes the total correlation in the initial entangled state, into  $\rho_{\sigma ,\mathbf{q} } $ and $\rho_{\sigma ,\mathbf{-q} } $. However, there is no such an identity for the logarithmic negativity, hence Schwinger effect does not redistribute quantum entanglement. The reason is that the qubit is not  coupled with the electric field. In the fermion model studied previously~\cite{lidaishi}, even when reduced to the   problem of an uncoupled qubit and the fermionic  mode coupled with the electric field,  redistribution exists both in  mutual information and in logarithmic negativity. See Eqs.~(38-39) in Ref.~~\cite{lidaishi}, where the fermionic mode coupled with the electric field is denoted as $\mathbf{p}$, and the uncoupled mode $\mathbf{q}$ is equivalent to a qubit.

\section{Effect of a pulsed electric field   \label{pulse} }

Now we investigate the effect of a pulsed electric field. Consider a Sauter-type electric field $E(t) = E_0 \mathrm{sech}^2(t /\tau)$ along   $z$ direction, where $\tau$ is the width of the pulsed electric field~\cite{sauter}. The gauge potential $A_\mu$ can be chosen as
\begin{equation}   A_\mu = \left( 0,0,0, -E_0\tau \tanh \left(\frac{t}{\tau }\right) \right),   \end{equation}
for which the Bogoliubov transformation yields~\cite{kim,Ebadi}
\begin{equation}  | \alpha_\mathbf{k} |^2  = \frac{ \cosh [ \pi \tau ( \omega^{\mathrm{out}}_k +  \omega ^{ \mathrm{in}}_k ) ] + \cosh ( 2 \pi \lambda ) } {2 \sinh ( \pi \tau  \omega ^{ \mathrm{in}}_k )\sinh \ ( \pi \tau \omega ^{\mathrm{out}}_k )}  ,   \label{pebalpha}    \end{equation}
\begin{equation}  | \beta_\mathbf{k} |^2  = \frac{ \cosh [ \pi \tau ( \omega^{\mathrm{out}}_k -  \omega ^{ \mathrm{in}}_k ) ] + \cosh ( 2 \pi \lambda ) } {2 \sinh ( \pi \tau  \omega ^{ \mathrm{in}}_k )\sinh \ ( \pi \tau \omega ^{\mathrm{out}}_k )}  ,  \label{pebbeta}     \end{equation}
where
\begin{equation}  \lambda = \sqrt{(eE_0\tau^2)^2 - \frac{1}{4}}  ,   \end{equation}
\begin{equation}  \omega^{\mathrm{in}}_k = \sqrt{(k_z + eE_0 \tau)^2 + k^2_\bot + m^2}   , \end{equation}
\begin{equation}  \omega^{\mathrm{out}}_k = \sqrt{(k_z - eE_0 \tau)^2 + k^2_\bot + m^2}   .  \end{equation}
As $\tau  \to 0$,   $E(t) \to 0$, then $| \alpha_\mathbf{k} |^2 \to 1$ and $| \beta_\mathbf{k} |^2 \to 0$, reducing the problem to the case without the electric field. As $\tau  \to  + \infty $,  $E(t) \to E_0$, which means $| \alpha_\mathbf{k} |^2 $ and $| \beta_\mathbf{k} |^2 $ reduce to the values in the case of the constant electric field.

The analyses and calculations for   $\rho_{\sigma ,\mathbf{q}}$ and $\rho_{\sigma ,\mathbf{-q}}$ above for the case of a constant electric field can be applied to the pulsed electric field, but with $| \alpha_\mathbf{k} |^2$ and $| \beta_\mathbf{k} |^2 $ now given in Eqs.~(\ref{pebalpha}) and (\ref{pebbeta}). Hence the mutual information and logarithmic negativity now depend on not only $E_0$ but also $\tau$.

In parallel with the above study on a constant electric field, we first investigate the influence of a pulsed electric field on the entanglement and correlation between   $ \sigma $ and mode   $\mathbf{q}$.

\begin{figure}
\centering
\scalebox{0.65}{\includegraphics{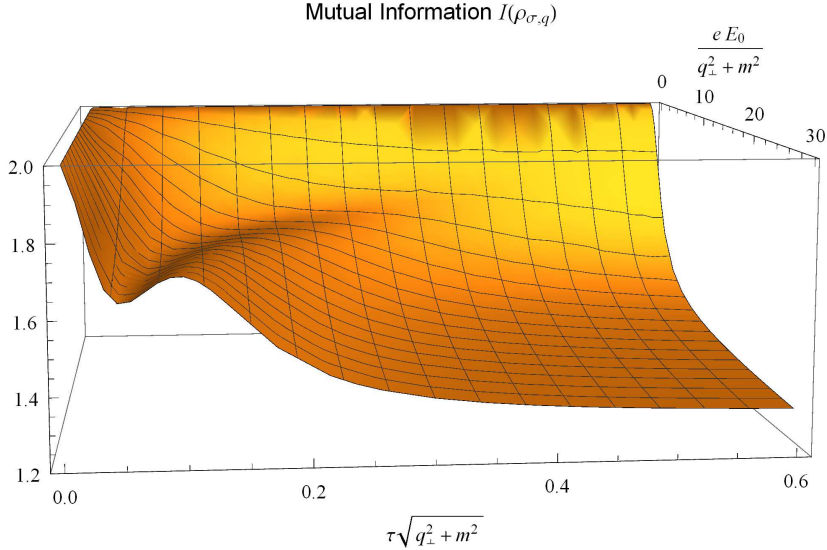}}
\caption{  The mutual information $I(\rho_{\sigma ,\mathbf{q}})$ as a function of  the dimensionless parameters $\frac{eE_0}{q_\perp^2 + m^2}$ and $\tau \sqrt{q_\perp^2+m^2 }$.   It is assumed  that  $ q_z=  \sqrt{q_\perp^2+m^2 }$,  $\varepsilon=1/ \sqrt{2}$.    \label{Fig4} }
\end{figure}

The influence of the pulsed electric field on the mutual information $I(\rho_{\sigma ,\mathbf{q}})$  is shown in
Fig.~\ref{Fig4}, which indicates its dependence  on the strength $E_0$ and the width $\tau$ of the pulsed electric field. For $E_0$ smaller than a certain value, with the increase of $\tau$, $I(\rho_{\sigma ,\mathbf{q}})$   monotonically decreases and approaches the   asymptotic value dependent on $E_0$ as $\tau \rightarrow \infty $. For $E_0$ larger than a certain value, with the increase of $\tau$, $I(\rho_{\sigma ,\mathbf{q}})$   first decreases to  a minimum  and then increases  to a maximum   before finally decreases and approaches asymptotically a value dependent on $E_0$ as $\tau \rightarrow \infty $.  The larger $E_0$ is, the smaller the values of $\tau$ corresponding to the minimum and  the maximum  of $I(\rho_{\sigma ,\mathbf{q}})$.  When $\tau$ is small, with the increase of $E_0$, $I(\rho_{\sigma ,\mathbf{q}})$     decreases to a minimum and then increases to a maximum, and finally decreases and approaches asymptotically  a certain value independent of $E_0$. For a given value of $\tau$, the variation trend of $I(\rho_{\sigma ,\mathbf{q}})$  with respect to $E_0$ is opposite to that of $I(\rho_{\sigma ,\mathbf{q}})$   with respect to $\tau$ for a given value of $E_0$.   Moreover,  When $\tau$ is smaller than a certain value,   the larger the values of $E_0$ corresponding to the minimum and the  maximum of $I(\rho_{\sigma ,\mathbf{q}})$.  When $\tau$ is larger than a certain value, $I(\rho_{\sigma ,\mathbf{q}})$ decreases monotonically  with the increase of $E_0$ and asymptotically approaches a value independent of $E_0$. As $\tau \rightarrow \infty$, the dependence of $I(\rho_{\sigma ,\mathbf{q}})$ on $E_0$  is the same as that the case of constant electric field  for $\varepsilon=1/ \sqrt{2}$, as shown in Fig.~\ref{Fig1}.

\begin{figure}
\centering
\scalebox{0.7}{\includegraphics{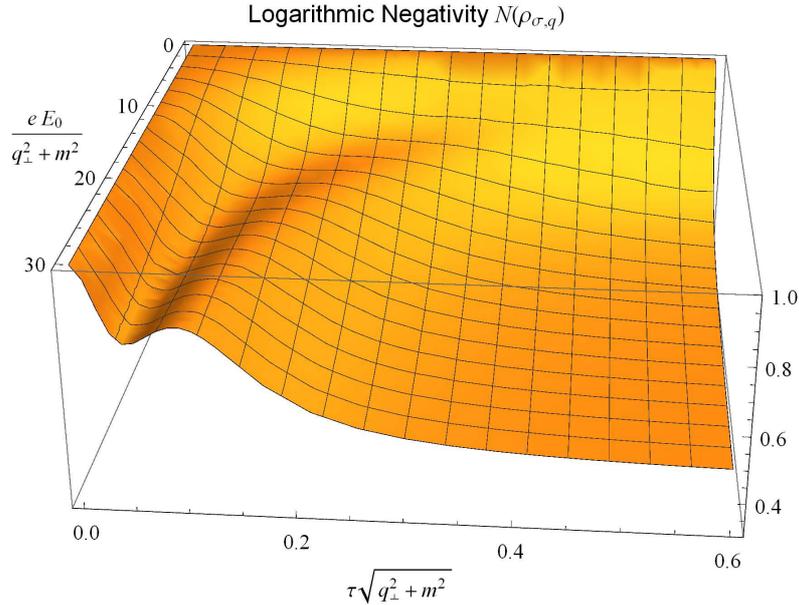}}
\caption{  The logarithmic negativity $N(\rho_{\sigma ,\mathbf{q}})$ as a function of  the dimensionless parameters $\frac{eE_0}{q_\perp^2 + m^2}$ and $\tau \sqrt{q_\perp^2+m^2 }$.   It is assumed  that  $ q_z=  \sqrt{q_\perp^2+m^2 }$,  $\varepsilon=1/ \sqrt{2}$.    \label{Fig5} }
\end{figure}

As shown in Fig.~\ref{Fig5}, the dependence of  $N(\rho_{\sigma ,\mathbf{q}})$ on   $E_0$ and $\tau$ is entirely similar to that of $I(\rho_{\sigma ,\mathbf{q}})$, but the values of $\tau$ and $E_0$   corresponding to the minima and maxima of $N(\rho_{\sigma ,\mathbf{q}})$ are different from those of  $I(\rho_{\sigma ,\mathbf{q}})$.    As $\tau \rightarrow \infty$ , the case of the constant electric field is also recovered, as shown in Fig.~\ref{Fig2} for $\varepsilon=1/ \sqrt{2}$.

\begin{figure}
\centering
\scalebox{0.65}{\includegraphics{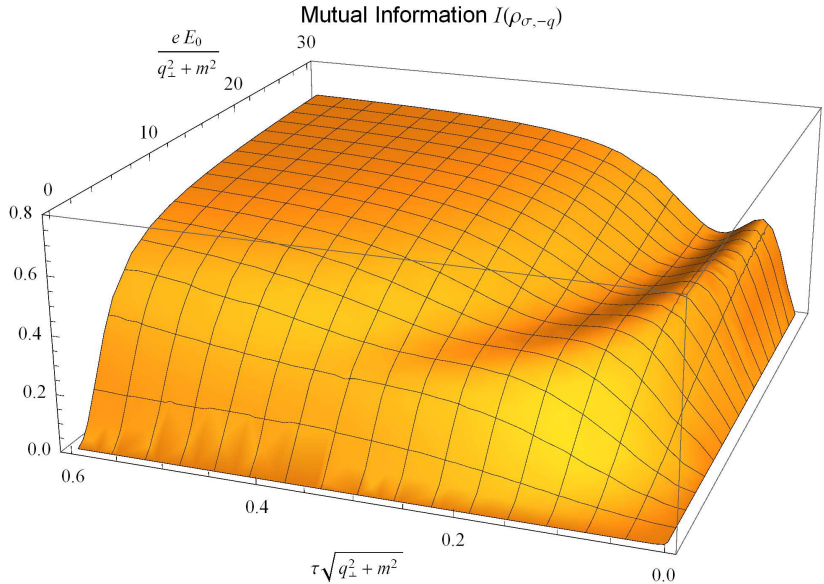}}
\caption{  The mutual information  $I(\rho_{\sigma ,\mathbf{-q}})$ as a function of  the dimensionless parameters $\frac{eE_0}{q_\perp^2 + m^2}$ and $\tau \sqrt{q_\perp^2+m^2 }$.   It is assumed  that  $ q_z=  \sqrt{q_\perp^2+m^2 }$,  $\varepsilon=1/ \sqrt{2}$.   \label{Fig6} }
\end{figure}

The correlation between $\sigma$ and  $\mathbf{-q}$ is exactly a complement of that between $\sigma$ and  $\mathbf{q}$, as indicated in Eq.~(\ref{compl}). Therefore the dependence of  $I(\rho_{\sigma ,\mathbf{-q}})$ on the pulsed electric field  is exactly opposite to that of $I(\rho_{\sigma ,\mathbf{-q}})$, as shown in Fig.~\ref{Fig6}.

\section{Summary and discussions  \label{summary} }

In this paper, we consider  a state in which a qubit is entangled with a bosonic mode. The  scalar field is coupled with  an electric field.

We have studied how the total correlation and the quantum entanglement in  $\rho_{\sigma ,\mathbf{q} } $ and $\rho_{\sigma ,\mathbf{-q} } $ depend on the electric field.   In the case of a  constant electric field, the mutual information $I(\rho_{\sigma ,\mathbf{ q} } ) $ decreases with the increase of the strength of the electric field and   approaches a certain nonvanishing value, implying that the total correlation between qubit $ \sigma  $ and mode $\mathbf{q}$ never vanishes. Similarly, the logarithmic negativity $N(\rho_{\sigma ,\mathbf{ q} } ) $ decreases with the increase of the strength of the electric field and approaches a certain nonvanishing value,  implying that the entanglement in $\rho_{\sigma ,\mathbf{q} } $ never vanishes  even if the strength of the electric field tends to infinity. For $\rho_{\sigma ,\mathbf{-q} } $,  the mutual information $I(\rho_{\sigma ,\mathbf{ -q} } ) $ increases with the increase of the strength of the electric field and asymptotically approaches a certain value. In fact, the sum of $I(\rho_{\sigma ,\mathbf{ q} } ) $ and  $I(\rho_{\sigma ,\mathbf{ -q} } ) $ is a constant determined by the initial state.

However no matter how strong  the electric field is, the logarithmic negativity $N(\rho_{\sigma ,\mathbf{ -q}} ) $ remains zero, i.e. $\sigma$ and mode $\mathbf{-q} $ remain unentangled, though Schwinger effect  does decrease the entanglement between $\sigma$ and mode $\mathbf{q}$.  We have also considered Schwinger effect of a  pulsed electric field, for which the pulse width plays a role.

The bosonic correlation and entanglement  an electric field are quite  different from those  of the fermionic entanglement ~\cite{lidaishi}. In the bosonic case,  with the increase of the electric field strength, the correlation (measured as mutual information)  and the entanglement (measured as mutual information) between the qubit and the original particle mode  decrease towards  non-zero asymptotic values, while the entanglement between the qubit and the   antiparticle mode remains zero  though the correlation increases towards an asymptotic value. In the fermionic case, with the increase of the electric field strength, the correlation and the entanglement   between the qubit and the original particle mode  decrease towards  zero, while both  the correlation and the entanglement   between the qubit and the antiparticle mode  increases towards values of those   between the qubit and the particle mode.

\vspace{0.5cm}

This work is supported by National Natural Science Foundation of China (Grant No. 11574054).

\end{document}